\documentclass[aps,amsmath,twocolumn,amssymb,floatfix,showpacs,superscriptaddress,footinbibliography]{revtex4-1}
\usepackage{amsmath}
\usepackage{mathtools}
\usepackage{braket}
\usepackage{bm}
\usepackage{graphicx}
\usepackage{amssymb}
\usepackage[dvipsnames]{xcolor}
\usepackage{multirow}
\usepackage{xfrac}
\usepackage{textcomp}
\usepackage{float}
\usepackage{enumerate}
\usepackage{subfigure}

\usepackage[colorlinks=true,linktoc=page,linkcolor=ForestGreen,citecolor=magenta,urlcolor=Purple]{hyperref}

\newcommand{\ie}{{\it i.e.,\,\,}}

\newcommand\bea{\begin{eqnarray}}
	\newcommand\eea{\end{eqnarray}}
\newcommand\beq{\begin{equation}}  
	\newcommand\eeq{\end{equation}}

\newcommand{\non}{\nonumber}  
\usepackage[normalem]{ulem}

\begin{document}
	

\title{Hierarchy of higher-order topological superconductor in three dimension}  

\author{Arnob Kumar Ghosh}
\email{arnob@iopb.res.in}
\affiliation{Institute of Physics, Sachivalaya Marg, Bhubaneswar-751005, India}
\affiliation{Homi Bhabha National Institute, Training School Complex, Anushakti Nagar, Mumbai 400094, India}
\author{Tanay Nag}
\email{tnag@physik.rwth-aachen.de}
\affiliation{Institut f\"ur Theorie der Statistischen Physik, RWTH Aachen University, 52056 Aachen, Germany}
\author{Arijit Saha}
\email{arijit@iopb.res.in}
\affiliation{Institute of Physics, Sachivalaya Marg, Bhubaneswar-751005, India}
\affiliation{Homi Bhabha National Institute, Training School Complex, Anushakti Nagar, Mumbai 400094, India}

\begin{abstract}
After exploring much on two-dimensional higher-order topological superconductors (HOTSCs) hosting Majorana corner modes (MCMs) only, we propose a simple fermionic model based on a three-dimensional topological insulator proximized with $s$-wave superconductor to realize Majorana hinge modes (MHMs) followed by MCMs under the  application of appropriate Wilson-Dirac perturbations. 
We interestingly find that the second-order topological superconductor, hosting MHMs, appears above a threshold value of the first type perturbation while the third-order topological superconducting phase, supporting MCMs, immediately arises incorporating infinitesimal perturbation of the second kind. Thus, a hierarchy of HOTSC phases can be realized in a single three-dimensional model. Additionally, the application of bulk magnetic field is found to be instrumental in manipulating the number of MHMs, leaving the number for MCMs unaltered. We analytically understand these above-mentioned numerical findings by resorting to the low energy model. We further characterize these topological phases with a distinct structure of the Wannier spectra.  From the practical point of view, we manifest quantized transport signatures of these higher-order modes. Finally, we construct Floquet engineering to generate the hierarchy of HOTSC phases by kicking the same perturbations as considered in their static counterpart.
\end{abstract}

\maketitle

\section{Introduction}

The advent of Majorana zero modes (MZMs) in topological superconductors (TSCs) prepares this field very relevant in the context of quantum information and topological quantum computations~\cite{Kitaev_2001,qi2011topological,hasan2010colloquium,das2012zero,Deng1557,Ivanov2001,nayak08}. Till date, there exist a variety of proposals based on heterostructures with spin-orbit coupling (SOC), such as one-dimensional (1D) nanowire with proximity induced $s$-wave superconductivity, 
that have been employed to engineer the MZMs~\cite{Fu2008,Sau2010,Lutchyn10,Hughes2010,Oreg2010}. 
In recent times, the higher-order topological (HOT) phases, harboring  boundary modes of lower dimension than their usual one, have been proposed with unconventional bulk-boundary correspondence. 
To pose a formal definition, an $n^{\rm th}$-order topological insulator~\cite{benalcazar2017,benalcazarprb2017,Song2017,Langbehn2017,schindler2018,Franca2018,wang2018higher,Ezawakagome,Roy2019,Trifunovic2019,Khalaf2018} or superconductor~\cite{Geier2018,Zhu2018,Liu2018,Yan2018,WangWeak2018,ZengPRL2019,Zhang2019,ZhangFe2019PRL,Volpez2019,YanPRB2019,Ghorashi2019,GhorashiPRL2020,Sumathi2020,Wu2020,jelena2020HOTSC,
BitanTSC2020,SongboPRR12020,SongboPRR22020,SongboPRB2020,kheirkhah2020vortex,PlekhanovArxiv2020,ApoorvTiwari2020,YanPRL2019,AhnPRL2020,luo2021higherorder2021,QWang2018} in $d$ dimensions is characterized by the existence of $n_c=(d-n)$-dimensional boundary modes. This bulk-boundary correspondence is further enriched for driven systems where non-trivial winding wave-functions in the temporal direction lead to dissipationless Floquet HOT insulator (FHOTI)~\cite{Bomantara2019,Nag19,YangPRL2019,Seshadri2019,Martin2019,Ghosh2020,Huang2020,HuPRL2020,YangPRR2020,Nag2020,ZhangYang2020,bhat2020equilibrium,GongPRBL2021,
chaudharyphononinduced2020} and superconductor (FHOTSC) phases~\cite{PlekhanovPRR2019,BomantaraPRB2020,RWBomantaraPRR2020,RWBomantaraPRR2020,ghosh2020floquet,ghosh2020floquet2}.

Very recently, a plethora of theoretical proposals have been put forward for realizing second-order topological superconductor (SOTSC) hosting zero-dimensional (0D) Majorana corner mode (MCM) in two-dimension (2D) and 1D Majorana hinge mode (MHM) in three-dimension (3D)~\cite{Geier2018,Zhu2018,Liu2018,Yan2018,WangWeak2018,ZengPRL2019,Zhang2019,ZhangFe2019PRL,Volpez2019,YanPRB2019,Ghorashi2019,GhorashiPRL2020,Sumathi2020,Wu2020,jelena2020HOTSC,BitanTSC2020,SongboPRR12020,SongboPRR22020,SongboPRB2020,kheirkhah2020vortex,PlekhanovArxiv2020,ApoorvTiwari2020,YanPRL2019,AhnPRL2020,luo2021higherorder2021,QWang2018}. However, the search for the 0D MCMs as a signature of third-order topological superconductor (TOTSC) in 3D is still in its infancy~\cite{Khalaf2018,YanPRL2019,AhnPRL2020,luo2021higherorder2021,PhysRevX.10.041014,WuHighTchigherorder}. We note that the previous studies mostly rely on unconventional/odd parity superconductivity~\cite{YanPRL2019,ZhangFe2019PRL,PhysRevX.10.041014,WuHighTchigherorder}. Therefore, a fundamental question remains, which is whether the TOTSC phase can be perceived employing the conventional $s$-wave superconductivity that we intend to answer here. Motivated by the studies on FHOTIs and FHOTSCs in 2D, the other relevant question is how to engineer the FHOTSC phases by periodically driving the appropriate perturbations in 3D~\cite{Bomantara2019,Nag19,YangPRL2019,Seshadri2019,Martin2019,PlekhanovPRR2019,Ghosh2020,Huang2020,HuPRL2020,BomantaraPRB2020,YangPRR2020,Nag2020,ZhangYang2020,
bhat2020equilibrium,GongPRBL2021,chaudharyphononinduced2020,RWBomantaraPRR2020,ghosh2020floquet,ghosh2020floquet2}. From the application point of view, the SOTSC (2D) and TOTSC (3D), harboring 
0D MCMs, can become a more suitable candidate for topological quantum computation compared to the other one-dimensional (1D) nanowire models where MZMs have been realized~\cite{Lutchyn10,Oreg2010}. In 1D wire networks, one has to engineer a $\rm T$-junction for the braiding of MZMs~\cite{Alicea2011}. Although it is possible to exchange the MZMs strictly in 1D, but there might not be strong topological protection~\cite{Chiu_2015,Kornich2021PRL}. However, in case of higher-order topological superconductor (HOTSC), there exists specially separated localized 0D MCMs in 2D and 3D, which can provide a better platform for braiding of non-local MCMs. Thus, non-abelian statistics due to exchange of the MCMs can be naturally expected to become feasible in HOTSC systems~\cite{SongboPRR22020}. However, to the best of our knowledge, no proposal has been reported so far regarding the advantage of TOTSC (3D) compared to SOTSC (2D) as far as braiding of non-local MCMs is corcerned. Additionally, given the recent developements on 3D HOTI models~\cite{benalcazar2017, Nag2020}, it is worth investigating a theoretical model for TOTSC. However, the real materials, hosting SOTSC/TOTSC phases, are yet to be discovered and no experiment has been carried out so far, in this regard, to the best of our knowledge. On the other hand, given the experimental progress on realization of HOT phases in solid-state systems~\cite{schindler2018higher,Experiment3DHOTI.VanDerWaals} and meta-materials~\cite{serra2018observation,xue2019acoustic,ni2019observation,imhof2018topolectrical,experimentFloquetHOTI,Experiment3DHOTI.aSonicCrystals}, we believe that our quests are very much timely and authentic as far as the theoretical advancement of the HOT field is concerned.

\begin{figure}
	\centering
	\subfigure{\includegraphics[width=0.35\textwidth]{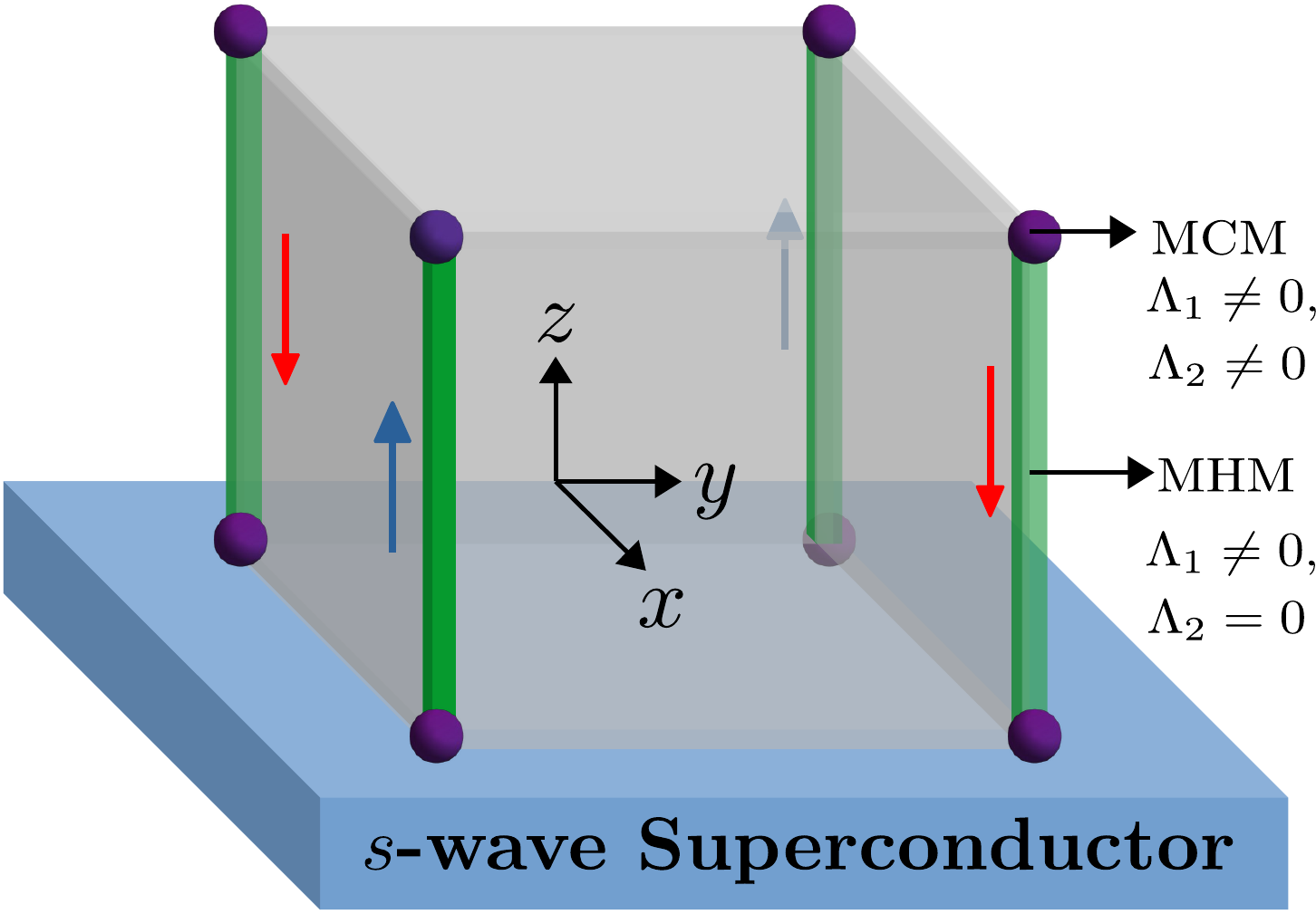}}
	\caption{(Color online) Schematic of our setup is demonstrated in which a cubic 3D TI (grey) is placed in close proximity to a bulk $s$-wave superconductor (light blue, light grey). 
	When $\Lambda_1\neq 0$ but $\Lambda_2=0$, MHMs are present as marked by the green (grey) line. For $\Lambda_{1,2} \neq 0$, MCMs appear as depicted by the purple (dark grey) dots. Blue and red arrows denote the propagation direction of MHMs.
	}
	\label{Schematics}
\end{figure}
\begin{figure}
	\centering
	\subfigure{\includegraphics[width=0.48\textwidth]{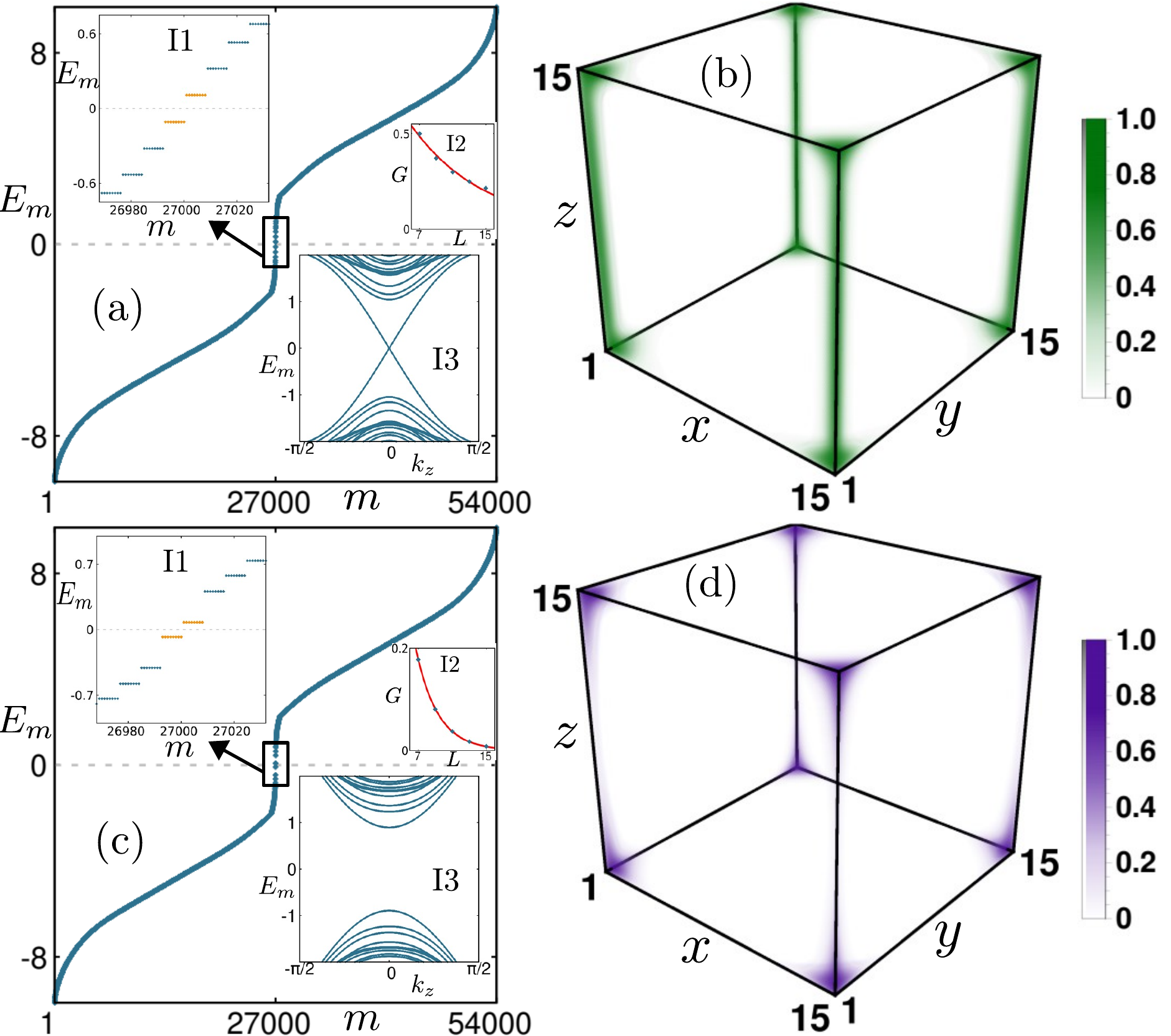}}
	\caption{(Color online) (a)~Eigenvalue spectrum $E_m$ of the Hamiltonian (Eq.(\ref{Ham})) for the SOTSC with $\Lambda_1=1.0$ and $\Lambda_2=0.0$, under OBC along all three directions, is shown as a function of state index $m$. The eigenvalue spectra close to $E_m=0$ is shown in the inset I1. Due to finite size effect, there exists a finite gap between the zero-energy modes. Although this gap, $G$ diminishes exponentially with increasing the system size~($G\sim a_1 \exp(-b_1 L)$, where $a_1=1.0634$, $b_1=0.1119$; with $L$ being the system-size in one direction.) as shown in the inset I2. In the inset I3, we depict the eigenvalue spectrum for the same Hamiltonian, but considering a rod geometry featuring the dispersive nature of the MHMs. (b)~The LDOS, associated with the MZMs appearing at $E_m =0$ for SOTSC, clearly establishes the existence of MHMs along $z$ direction at the interfaces of $xz$ and $yz$ surfaces of the cube. (c)~We repeat (a) with $\Lambda_1=1.0$ and $\Lambda_2=1.0$ for TOTSC. The zoom-in spectra near $E_m=0$ is shown in the inset I1.
	Here, the finite-size gap $G\sim a_2 \exp(-b_2 L)$ with $a_2=2.7413$ and $b_2=0.3908$ is depicted in the inset I2. The inset I3 indicates that the eigenvalue spectrum in the rod geomrtry is gapped due to the inclusion of $\Lambda_2\neq 0$. (d)~The corresponding LDOS structure for MZMs demonstrates very strong corner localization namely, MCMs in TOTSC. The value of the other parameters are chosen to be $m=2.0,t=\lambda=1.0, \Delta=0.4$, $h_x=0.0$. 
	}
	\label{SecondDOS}
\end{figure}

In this article, we come up with a model to systematically realize MHMs in SOTSC and MCMs in TOTSC, starting from a 3D topological insulator (TI) proximized by $s$-wave superconductivity, 
through applying appropriate perturbations (see Fig.~\ref{Schematics} and Fig.~\ref{SecondDOS}) in bulk. The effect of a bulk magnetic field remarkably imprints its effect on the SOTSC while the TOTSC remains unaffected. We topologically characterize such phases by investigating the Wannier spectra (WS) (see Fig.~\ref{Wannier}). We also exhibit the signature of the SOTSC phase by calculating the differential conductance through the MHMs following a system-lead setup (see Fig.~\ref{Wannier}). This further enriches the experimental relevance of our work. Moreover, we extend our analysis to selectively generate FHOTSC phases starting from a trivial superconducting phase in 3D (see Fig.~\ref{FloquetLDOS}).

The remainder of the paper is organized as follows. In Sec.~\ref{Sec:II}, we introduce and motivate our model along with the discussion of various phases available for the system. Emergence of SOTSC and TOTSC is discussed in Sec.~\ref{Sec:III} and the detail derivation of the surface Hamiltonian, hinge Hamiltonian, and corner mode solutions are provided in Appendices~\ref{App:A}, \ref{App:B}, and \ref{App:C}, respectively. Sec.~\ref{Sec:IV} is devoted to the topological characterization of MHMs and MCMs. In Sec.~\ref{Sec:V}, we provide the transport calculation for SOTSC and the lattice model setup used for our transport calculation is illustrated in Appendix~\ref{App:D}. Floquet generation of HOTSC with a specific form of the driving protocol is briefly discussed in Sec.~\ref{Sec:VI}. Finally, we summarize and conclude our paper in Sec.~\ref{Sec:VII}.

\section{Model and Motivation}\label{Sec:II}

To begin with, we introduce a Bogoliubov-de Gennes (BdG) Hamiltonian on a cubic lattice incorporating $s$-wave superconductivity induced in 3D TI~\cite{zhang2009topological,slager14} via the proximity effect
\begin{equation}
	H_{0}(\bm k)=
	\begin{pmatrix}
		H_{\rm TI} (\bm k)-\mu & \Delta \\
		\Delta^* & \mu-\mathcal{T}^{-1} H_{\rm TI} (\bm {-k}) \mathcal{T}\   \\
	\end{pmatrix}\ ,
	\label{eq1_TSC}
\end{equation}
with TI model $H_{\rm TI} (\bm k) = 2 \lambda \sum_{j=x,y,z} \sin k_j \mu_x\sigma_x s_j + \left[(m_0-6t)+2t \sum_{j=x,y,z}\cos k_j \right] \mu_x\sigma_z$ where, $t$ ($\lambda$) represents the nearest-neighbor hopping (SOC) strength, $\Delta=\Delta_0$ is the $s$-wave superconducting pairing gap induced via the proximity effect, $m_0$ is the crystal-field splitting energy and $\mu$ is the chemical potential. The on-site mass term $m_0$ plays a very crucial role in the band inversion phenomena while combined with $\cos k_j$ terms for the TI model \cite{hasan2010colloquium,bernevig2006quantum}. The Pauli matrices ${\bm \mu}$, ${\bm \sigma}$, and ${\bm s}$ act on sub-lattice $(A,B)$, orbital $(\alpha,\beta)$, and spin $(\uparrow, \downarrow)$  degrees of freedom respectively.

Note that, $H_{\rm TI} (\bm k)$ supports strong TI phase ($\mathcal {Z}_{2}$ class) with bulk band inversion taking place at ${\bm \Gamma}=(0,0,0)$ point of the Brillouin zone for $0<m_0/t<4$ \cite{slager14}. For $4<m_0/t<8$, the band inversion occurs at ${\bm M}=\{(0,\pi,\pi),~(\pi,0,\pi),~(\pi,\pi,0)\}$ points and the model supports weak TI phase. On the other hand, band inversion takes place at ${\bm R}=(\pi,\pi,\pi)$ point in the strong TI phase for $8<m_0/t<12$ while the TI model becomes trivially gapped for $m_0/t>12$.  
All these above TI phases are first order exhibiting gapless surface states that are protected by the time-reversal symmetry (TRS) $\mathcal{T} = i s_y \mathcal{K} $ where, $\mathcal{K}$ denotes the complex-conjugation operator. We restrict ourselves to the strong topological phase to start, with $m_0/t=2$ throughout the manuscript, unless mentioned otherwise.

Very recently, second-order TI (SOTI) phases in 2D are shown to be elevated to third-order TI (TOTI) phases in 3D~\cite{benalcazar2017,Nag2020}. In particular, starting from first order TI (FOTI) phases, the ladder of HOTI \ie SOTI and TOTI phases can be engineered consistently via discrete symmetry breaking perturbations~\cite{Nag2020}. Below we elaborate them individually.
When one incorporates the HOT mass term $V_1 \mu_x \sigma_y$ with $V_1=\sqrt{3} \Lambda_1 \left(\cos k_x-\cos k_y \right)$, it gaps out the 2D surface modes of FOTI exhibiting four intersections between $xz$ and $yz$ surfaces gapless as $V_1$ vanishes along $k_x=\pm k_y$. The $C_4$ symmetry breaking Wilson-Dirac mass $V_1$ changes its sign between the above surfaces leading to the SOTI with gapless 1D hinge modes along $z$-direction: $H_{\rm SOTI} =H_{\rm TI} + V_1 \mu_x \sigma_y$ \cite{Nag2020}. 
Interestingly, the SOTI mass term $V_1 \mu_x \sigma_y$ breaks TRS $\mathcal{T}$, however, chiral hinge modes are preserved by $C_4 \mathcal{T}$ symmetry.
The introduction of another Wilson-Dirac mass term $V_2 \mu_z$ with $V_2=\Lambda_2 \left(2\cos k_z-\cos k_x-\cos k_y \right)$  would result in a TOTI with zero-energy corner modes residing only at the eight corners of the cubic system:  $H_{\rm TOTI}=  H_{\rm SOTI}+ V_2 \mu_z$ \cite{Nag2020}. This is due to the fact that $V_2$ gaps out the hinge modes while it vanishes over eight body-diagonals $\pm k_x=\pm k_y =\pm k_z$. We emphasize that these zero-energy HOT modes respect unitary chiral and anti-unitary particle hole symmetry~\cite{Nag2020}. We additionally note that, as long as the lattice termination has to be compatible with the four fold rotation symmetry, the hinge/corner modes continue to exist~\cite{trifunovic2021higher}. 

Inspired by the recent theoretical study on the 3D HOTI phases as discussed above~\cite{Nag2020}, we here present a new 3D model that allows us to explore the SOTSC and TOTSC phases systematically. We note that unlike the FOTI, there is no first-order TSC phase 
to start with as the BdG Hamiltonian (Eq.(\ref{eq1_TSC})) becomes trivially gapped out by the superconducting pairing gap $\Delta_0$. We assume a constant superconducting gap over the entire sample without taking into account the microscopic description of this proximity induced gap~\cite{faraei17}.
Inspired by the fact that in 2D, a trivial $s$-wave superconductor is proposed to host MCMs in the presence of a magnetic field~\cite{Wu2020}, we also consider the TRS breaking magnetic field $h_x s_x$ with $H_{\rm TI}$ in 3D. At the outset, we propose a generic Hamiltonian, combing $H_{0}(\bm k)$ (Eq.(\ref{eq1_TSC})) with the relevant perturbations $V_1$, $V_2$, and $h_x$ that can host the HOTSC phases, as follows~\cite{benalcazar2017,Nag2020}
\begin{eqnarray}\label{Ham}
	H(\bm k)&=& 2 \lambda \sum_{j=1}^3 \sin k_j \Gamma_j
	+\left[(m_0-6t)+2t \sum_{j=1}^3 \cos k_j \right] \Gamma_4 \nonumber \\
	&&+\Delta_0 \Gamma_5 + V_1 \Gamma_6 + 
	V_2  \Gamma_7  +h_x \Gamma_8 = {\bm N}(\bm k)\cdot {\bm \Gamma}\ ,
\end{eqnarray}
with the convention $k_{1,2,3}=k_{x,y,z}$, ${\bm N}(\bm k)=\left( N_1(\bm k), \cdots, N_8(\bm k) \right)$ and ${\bm \Gamma }= ( \Gamma_1, \cdots, \Gamma_8)$. Here $\Gamma$'s are $16 \times 16$ matrices: $\Gamma_1=\mu_x\sigma_x s_x \tau_z$, $\Gamma_2=\mu_x \sigma_x s_y\tau_z$,  $\Gamma_3=\mu_x\sigma_x s_z \tau_z$, $ \Gamma_4=\mu_x\sigma_z \tau_z$, $\Gamma_5=  \tau_x$, 
$\Gamma_6=\mu_x \sigma_y$, $\Gamma_7=\mu_z \tau_z$ and $\Gamma_8=s_x$. Here, ${\bm \tau}$ acts on the particle-hole subspace. This Hamiltonian (Eq.(\ref{Ham})) breaks TRS but preserves the particle-hole symmetry (PHS) $\mathcal{C}= s_y\tau_y \mathcal{K}$. We now analyze individual situation by considering $V_1$ perturbation only (\ie $V_2=0$) and then $V_1$, $V_2$ perturbations together. These scenarios allow us to investigate the cascade of HOTSC phases in 3D. The corresponding real space tight-binding verion of our HOTSC model (Eq.(\ref{Ham})) is demonstrated in 
Eq.(\ref{real_space}) with the on-site superconducting pairing gap $\Delta_0$ term.
For sake of simplicity, we consider $\mu=0$ throughout our analysis. However, the chemical potential can also be finite (inside the bulk gap) in order to realize HOT modes.
This allows us to probe the influence of discrete symmetry breaking mass perturbation instead of the chemical potential driven transitions.

We further emphasize that the above TSC model (Eq.(\ref{Ham})) is not directly connected to any material platform, rather can be thought of as a theoretical framework to generate the hierarchy of higher-order Majorana modes in 3D. Interestingly, the underlying HOTI model can be formulated in several ways by considering different representations of $8 \times 8$ Hermitian 
matrices~\cite{Nag2020}. As a result, there exist a lot of freedom to choose other  BdG compatible representations of $16 \times 16$ $\Gamma$ matrices in order to formulate TSC model that hosts MCMs. With the construction of the HOTSC model in general, we believe that our model could turn out to be useful in explaining future HOTSC findings from real materials perspective with broken TRS. Note that, TI/ TSC model can become block-diagonal in certain representation for some choice of degrees of freedom such as, orbital, spin, sub-lattice, etc., the anti-unitary symmetry plays a crucial role in confining boundary modes at zero-energy~\cite{BitanTSC2020}.

\section{Emergence of second and third-order topological superconductor}\label{Sec:III}
To explore the appearance of 1D MHMs propagating along $z$ direction at the intersection of $xz$ and $yz$ surfaces, we first consider $V_1 \ne 0$ and $V_2=0$. We numerically diagonalize the Hamiltonian (Eq.(\ref{Ham})), considering open boundary condition (OBC) in all three directions, to show the zero-energy states and corresponding local density of states (LDOS) in Fig.~\ref{SecondDOS} (a) and (b), respectively. We next consider $V_1\ne0$ and $V_2 \ne 0$ in Hamiltonian (Eq.(\ref{Ham})) to investigate the MCMs in TOTSC phase with the notion that $V_2$ vanishes along body diagonals. Our numerical findings clearly depict that the MHMs become gapped out, leaving only zero-energy mid-gap TOTSC states as shown in eigenvalue spectra in Fig.~\ref{SecondDOS}(c) while the associated LDOS demonstrates sharp corner localization in Fig.~\ref{SecondDOS}(d). The above results are presented for $h_x=0$. However, we note that these observations remain qualitatively unaltered for finite value of $h_x$ (see the text below for more details from an analytical viewpoint). The zoom-in spectra near $E_m=0$ and the finite size gap analysis are respectively depicted in insets I1 and I2 of Fig.~\ref{SecondDOS}(a) (Fig. \ref{SecondDOS}(c)) for MHMs (MCMs).
We stress that chiral MHMs are dispersive along $z$ direction as clearly observed when the SOTSC Hamiltonian is studied employing a rod geometry with $k_z$ as one of the good quantum number (see inset I3 of Fig.~\ref{SecondDOS}(a)). On the other hand, zero-energy MCMs cannot be captured from ${\bm k}$-space dispersion as they always appear to be gapped (see inset I3 of Fig.~\ref{SecondDOS}(c)) and localized at the corners. One can find 16 zero-energy states $E_m \simeq 0$ in the second and third order phases \ie there exist four (two) gapless MHMs (zero-energy MCMs) per hinge (coner). These MZMs are not Kramer's pairs as TRS symmetry is explicitly broken in the higher order phases by the Wilson-Dirac mass terms.
It is imperative to mention here 
that all the parameters/quantities having the dimension of energy are scaled by the hopping strength $t$. The lattice spacing is set to be unity throughout our analysis.

\begin{figure}
	\centering
	\subfigure{\includegraphics[width=0.48\textwidth]{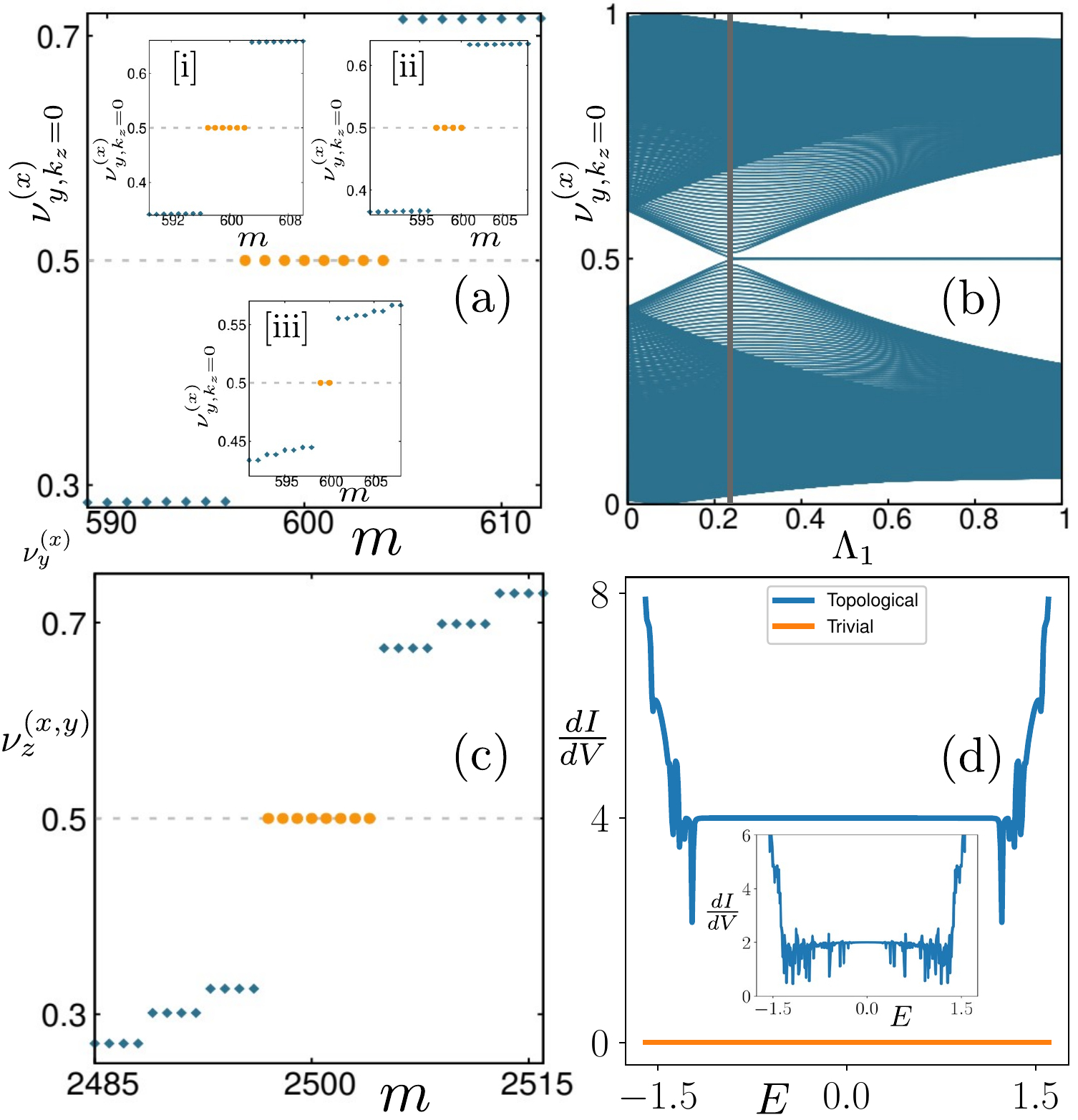}}
	\caption{(Color online) (a)~The WS $\nu^{(x)}_{y,k_z=0}$, computed using Eq.(\ref{WS}), is illustrated for SOTSC with $\Lambda_1=0.8$ and $\Lambda_2=0.0$, as a function of the state-index with    OBC along $x$ direction and PBC along $y$ and $z$ directions. One can observe eight eigenvalues that appear at $0.5$ corresponding to $16$ MHMs for $m=2.0$ and $h_x=0.0$. In inset [i], [ii] and [iii] we  show six, four and two eigenvalues at $0.5$ in WS for $12$, $8$ and $4$ MHMs respectively when $(\Lambda_1, h_x)=(0.8, 1.2)$, $(0.8, 2.0)$ and $(0.1, -0.5)$ respectively. (b)~The WS 
	$\nu^{(x)}_{y,k_z=0}$ is shown as a function of $\Lambda_1$ where the second order topological phase transition takes place for $M_{\Lambda_1} >M^T_{\Lambda_1}=M_{\Delta}$. (c)~The WS 
	$\nu^{(x,y)}_{z}$ for TOTSC is depicted as a function of the state-index considering OBC along $x$ and $y$ directions and PBC along $z$ direction. We choose $\Lambda_1=1.0$, $\Lambda_2=1.0$,  
	$m=2.0$ and $h_x=0.0$. Here, the eight eigenvalues at $0.5$ correspond to the $16$ MCMs. (d)~Differential conductance $\frac{dI}{dV}$ (in the unit of $\frac{e^2}{\hbar}$) is shown as a function  
	of the incident electron energy $E$ for SOTSC (trivial SC) phase when $m=2.0$ ($m=20.0$), $\Lambda_1=1.0$, $\Lambda_2=0.0$ and $h_{x}=0.0$. Inset represents the case where 
	we have two MHMs per hinge. We choose the parameters for this case as $(\Lambda_1,h_x)=(0.8,2.0)$ and $\Lambda_2=0.0$.
	}
	\label{Wannier}
\end{figure}

We anchor the above findings with the low-energy effective Hamiltonian where we rigorously investigate the effect of $h_x$  on the number of MHMs. Note that, $h_x$ is treated perturbatively with respect to the bulk gap of the underlying TI while deriving the low-energy Hamiltonians.
We procure the surface Hamiltonian 
$H^S_{ij}$ for $ij$ surface in the projected basis (See Appendix~\ref{App:A}), imposing OBC for the remaining $k$ direction in Hamiltonian (Eq.(\ref{Ham})),  as~\cite{ghosh2020floquet2}
\begin{eqnarray}
H^{\rm S}_{xy}&=&-2 \lambda k_x \sigma_x s_y \tau_z +2 \lambda k_y \sigma_x s_x \tau_z + M_{\Delta}  \tau_x - 2M_{\Lambda_2} \sigma_z \tau_z, \non \\
H^{\rm S}_{yz}&=&2 \lambda k_y \sigma_x s_x \tau_z  +2 \lambda k_z \sigma_x s_y \tau_z + M_{\Delta}  \tau_x -M_{\Lambda_1} \sigma_x s_z   \non \\
&&+h_x s_z + M_{\Lambda_2} \sigma_z \tau_z, \non \\
H^{\rm S}_{xz}&=&-2 \lambda k_x \sigma_x s_x \tau_z  +2 \lambda k_z \sigma_x s_y \tau_z + M_{\Delta} \tau_x +M_{\Lambda_1} \sigma_x s_z   \non \\
&&+ M_{\Lambda_2} \sigma_z \tau_z,
\label{surface}
\end{eqnarray}
with, $M_{\Delta}=\Delta_0$, $M_{\Lambda_1}=\frac{\sqrt{3} m_0\Lambda_1}{2t}$ and $M_{\Lambda_2}=\frac{m_0 \Lambda_2}{2t}$. We would like to stress on the fact that the bulk magnetic field in 3D has finite projections on the 2D surface as can be seen from the aforementioned surface Hamiltonians. Now focusing on $xz$ and $yz$ surface Hamiltonians, we find a set of common terms that would participate identically to build up an effective gap in the above two surfaces (See Appendix~\ref{App:A} for detail derivations of the surface Hamiltonian). Let us first analyze the MHMs from surface Hamiltonian (Eq.(\ref{surface})) considering $\Lambda_2=0$ and $h_x=0$. The term corresponding to Wilson-Dirac mass, $M_{\Lambda_1}$ changes its sign between the above two surfaces, resulting in $16$ gapless MHMs for $M_{\Lambda_1}>M_{\Delta}$~\cite{ghosh2020floquet2} (see Figs.~\ref{SecondDOS}(a) and (b)). This is due to the fact that all $8$  gap factors, obtained by considering $\tau_x,~\sigma_x,~s_z= \pm 1$, change their sign accordingly. Therefore, unlike SOTI that immediately arises for any non-zero values of $M_{\Lambda_1} $, the SOTSC phase can only emerge above a threshold value of $M^T_{\Lambda_1}$ such that $M_{\Lambda_1}>M^T_{\Lambda_1} = M_{\Delta}$. After introducing the magnetic field with $h_x > 0$, $16$ MHMs continue to exist as long as $h_x < M_{\Lambda_1}- M_{\Delta}$. For $M_{\Lambda_1}- M_{\Delta} < h_x < M_{\Lambda_1}+ M_{\Delta}$, there are $6$ gap factors that reverse their sign between the above two surfaces leading to $12$ MHMs. On the other hand, for $h_x > M_{\Lambda_1}+ M_{\Delta}$, one can find $8$ MHMs in accordance with $4$ sign-changing gap factors between the above surfaces. Note that, one can surprisingly obtain $4$ MHMs with only $2$ sign-changing gap factors for $h_x <0$ (direction of the magnetic field is reversed) and $M_{\Lambda_1} < M_{\Delta}$ such that 
$M_{\Delta}-M_{\Lambda_1}<\lvert h_x \rvert<(M_{\Lambda_1} + M_{\Delta})$. This refers to the fact that the magnetic field can in principle alter $M^T_{\Lambda_1}$ as compared to the no magnetic field case. The above discussion is useful to understand the topological characterization of various SOTSC phases as depicted in Fig.~\ref{Wannier}. 

We now analytically explore the hinge Hamiltonian $H^H_{i,ij}$ for $i^{\rm th}$ hinge, obtained by imposing OBC in $j^{\rm th}$ direction on the surface Hamiltonian (Eq.(\ref{surface})), 
to investigate the MCMs in TOTSC phase (See Appendix~\ref{App:B} for details). The hinge Hamiltonian in the projected basis are as follows:
\small
\begin{eqnarray}
H_{x,xy}^H&=&2 \lambda k_x \tau_y - 2 M_{\Lambda_2} s_x \tau_x, \ H_{y,yz}^H=-2 \lambda k_y s_z \tau_z + M_{\Lambda_2} \tau_x, \non \\
H_{z,xz}^H&=&-2 \lambda k_z s_z \tau_z + M_{\Lambda_2}  \tau_x .
\label{hinge1}  
\end{eqnarray}
\normalsize
Note that, such a set of hinge Hamiltonian predicts the number of MCMs at any given corner with $M_{\Lambda_2} \ne 0$ causing the MHMs to be gapped out. 
The relative signs of gap factors, obtained by considering $s_x$, $\tau_x=\pm 1$, between any two hinge Hamiltonians change only for $s_x=+1$ block referring to the fact that each corner can host two MZMs in principle. Therefore, there exist 16 MCMs in the TOTSC altogether. Interestingly, the magnetic field does not appear in the hinge Hamiltonian and gap factors are insensitive to $h_x$ irrespective of its strength. However, the perturbation scheme breaks down if the strength of the magnetic field is arbitrarily large exceeding the bulk gap of the TI.
The bulk magnetic field in 3D does not have any projections on 1D hinge unlike the finite surface projection as given in Eq.(\ref{surface}).
One can consider adhoc surface magnetic field instead of incorporating it in bulk (Eq.(\ref{Ham})) such that the gap factors in the hinge Hamiltonian can be tuned with $h_x$~\cite{future_study}.  
 
To complete our study further, we investigate the wave-functions of MCMs for the simple case $h_x=0$~(See Appendix~\ref{App:C} for details).  After few lines of algebra while writing the surface Hamiltonian in terms of hinge Hamiltonian $H^S_{yz}= H_0 + H^H_{z,yz}$ with $\Lambda_2 =0$, the wave-function of a given hinge from $H_0$ can be found as $\Phi_{\rm MHM}\sim \sum^{4}_{n=1} N_n e^{-\alpha x} e^{-\beta_n y}  e^{i k_z z} \phi_n$. Here, $\alpha= \lambda/t$, $ \beta_1=\beta_2=(M_{\Lambda_1}- M_{\Delta})/2 \lambda$,  $\beta_3 = \beta_4=(M_{\Lambda_1}+ M_{\Delta})/2 \lambda$ and $\phi_n$ represents the spinor part. The  wave-function at a given corner, obtained from the hinge Hamiltonian $H^H_{z,yz}$ with OBC along $z$ hinge  and $\Lambda_2\ne 0$, can be found as $\Phi_{\rm MCM}\sim \sum^{2}_{n=1} N'_n e^{-\alpha x} e^{-\beta_1 y} e^{-\gamma_n z} \phi'_n $ with $\gamma_1= M_{\Lambda_2}/ 2 \lambda$ and $\gamma_2= M_{\Lambda_2}/ \lambda$. The localization length of MCMs and MHMs varies in   different directions and can in principle depend on $h_x$, if it is applied on the surface. Our analytical findings thus confirm the numerical observations for MHMs and MCMs as depicted in Figs.~\ref{SecondDOS}(b) and (d), respectively.

\section{Topological characterization}\label{Sec:IV}

Having established the HOTSC phases analytically, we now characterize them by investigating their respective Wannier spectra (WS). We employ periodic boundary condition (PBC) along two directions 
$y$ and $z$, and OBC along $x$ direction, to compute WS for the SOTSC phase. We construct the Wilson loop operator~\cite{benalcazarprb2017,ghosh2020floquet,ghosh2020floquet2} as follows
\begin{equation}
	{\mathcal W}^{(x)}_y=F^{(x)}_{y,k_y + (N_i -1) \Delta k_y,k_z } \cdots F^{(x)}_{y,k_y + \Delta k_y,k_z } F^{(x)}_{y,k_y,k_z}\ ,
	\label{WS} 
\end{equation} 
with $\left[F^{(x)}_{y,k_y,k_z}\right]_{mn}=\langle \phi^{(x)}_{n, k_y + \Delta k_y,k_z} | \phi^{(x)}_{m,k_y,k_z} \rangle$, where $\Delta k_i= 2\pi /N_i$ ($N_i$ being the number of discrete points considered inside the Brillouin zone (BZ) along $k_i$) and $|\phi^{(x)}_{m,k_y,k_z} \rangle $ is the $m^{\rm th}$ occupied state of the Hamiltonian (Eq.(\ref{Ham})). The corresponding Wannier Hamiltonian is given as ${\mathcal H}_{{\mathcal W}^{(x)}_{y}}= -i \ln {\mathcal W}^{(x)}_{y}$ whose eigenvalues $2\pi \nu^{(x)}_{y,k_z}$ correspond to the WS. We focus on $k_z=0$ point and show $\nu^{(x)}_{y,k_z=0}$ as a function of the state index $m$ in 
Fig.~\ref{Wannier}(a) when $h_x=0$. There exist eight eigenvalues at $0.5$ corresponding to an average of four MZMs to be present per hinge in the SOTSC phase. This corroborates with the $8$ 
sign changing gap factors in the low energy surface Hamiltonian (Eq.(\ref{surface})). By contrast, the WS of the trivial phase does not exhibit the eigenvalues at $0.5$.  The topological phase transition 
at $M^T_{\Lambda_1}=M_{\Delta}$, can thus be appropriately signalled by the feature of WS as depicted in Fig.~\ref{Wannier}(b). By tuning the magnetic field $h_x$, we obtain six, four, two eigenvalues 
at $0.5$ corresponding to $6$, $4$ and $2$ sign changing gap factors respectively as shown in the insets of Fig.~\ref{Wannier}(a) (see [i], [ii] and [iii], respectively).

Turning to the identification of the TOTSC phase with $M_{\Lambda_2}\ne 0$, we compute WS $\nu^{(x,y)}_z$, considering PBC along $z$ direction and OBC along $x$ and $y$ directions, as illustrated 
in Fig.~\ref{Wannier}(c). In the topological phase, we obtain eight eigenvalues at $0.5$, corresponding to an average of two MZMs to be present per corner. Application of infinitesimal bulk magnetic field $h_{x}$ does not alter the number of MCMs in this case as $h_{x}$ does not have any projections on 1D hinge.

\section{Transport signature of SOTSC phase}\label{Sec:V}
\begin{figure}[H]
	\centering
	\subfigure{\includegraphics[width=0.45\textwidth]{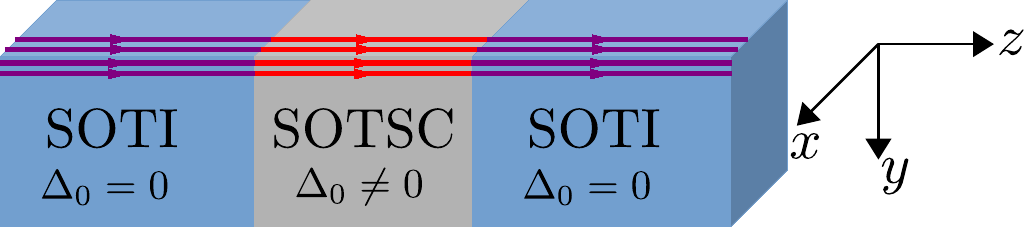}}
	\caption{(Color online) Schematic diagram of our transport setup that has been used to calculate the differential conductance for the MHMs. Hinge modes in the SOTI leads and in the middle SOTSC region are represented by purple and red lines respectively. When the number of modes in the lead matches with the number of modes in the central region, one obtains complete transmission of electrons via the hinge modes.}
	\label{TransportSchematics}
\end{figure}

The intriguing transport properties of helical Majorana edge modes are studied earlier~\cite{huangtansport2018,Litansport2020}. Here, we investigate the transport signature of propagating MHMs in 
SOTSC phase. In this purpose, we consider a two-terminal SOTI-SOTSC-SOTI setup (See Fig.~\ref{TransportSchematics} for schematics of the transport setup and Appendix~\ref{App:D} 
for the corresponding real space Hamiltonian). The incident current is injected from the left SOTI lead, propagates through the SOTSC, and output current $I$ is collected at the right SOTI lead while a 
potential difference 
\noindent
$e \left(V_L-V_R\right)\equiv eV$ is applied between the left (right) lead with voltage $V_L$ ($V_R$).  We choose this setup as the SOTI hosts $8$ gapless chiral electronic modes propagating along $z$-hinge while SOTSC (with $h_x =0$) harbors $16$ MHMs. Owing to the chiral nature of the MHMs, all the electronic modes from the left SOTI get transmitted to the right SOTI. The matching between the number of electronic modes (\ie 2 electronic hinge modes in SOTI) and its corresponding Majorana modes (\ie 4 MHMs in SOTSC) per hinge results in a complete transmission of the injected electronic modes while crossed Andreev reflection, normal electron reflection and Andreev reflection remain vanishingly small for this particular setup. The output current, $I$ for this setup is given by the Landauer-B\"{u}ttiker formula~\cite{landauer}
\begin{equation}\label{current}
I=\frac{e^2}{h} \mathcal{T}_{12} \left(V_L-V_R\right) \ ,
\end{equation}
where, $\mathcal{T}_{12}$ is the total normal electron transmission probability from the left lead to the right lead. From Eq.(\ref{current}) one obtain $\frac{dI}{dV} = \frac{e^2}{h} \mathcal{T}_{12}$ in the linear response regime. To obtain the signature of the MHMs, we calculate the differential conductance $\frac{dI}{dV}$ using KWANT~\cite{groth2014kwant} and depicted in Fig.~\ref{Wannier}(d).

Furthermore, the mid-gap MHMs exhibit quantized transport signature as long as the incident electron energy $E$ lies inside the bulk gap in the SOTSC phase. As on average, four gapless modes 
contribute per hinge, the $\frac{dI}{dV}$ exhibits quantized signal of $\frac{4e^{2}}{h}$ at $E=0$. Note that, the contribution in $\frac{dI}{dV}$, arising from electron transmission, vanishes when the central superconducting region becomes topologically trivial and does not support any gapless MHMs. Interestingly, by tuning $h_x$ in the SOTSC region 
(central region in Fig.~\ref{TransportSchematics}), there exists less number of MHMs in the SOTSC leading to a mismatch between the number of modes (\ie electronic modes and their corresponding Majorana modes) in the SOTI leads and the central SOTSC segment. Therefore, for completeness, in the inset of Fig.~\ref{Wannier}(d), we present the $\frac{dI}{dV}$ for the case when two gapless MHMs ($h_{x}\neq 0$) per hinge in SOTSC phase participate in transport. This refers to a  mismatch with respect to the number of corresponding electronic hinge mode present in the SOTI as discussed above. Note that, $\frac{dI}{dV}$ drops down to half of the earlier case \ie $\frac{2e^{2}}{h}$ at $E=0$. Turing to TOTSC, one can also attach a SOTI lead to one side of the TOTSC system to identify the transport signature of MCMs via an expected zero-bias peak in $\frac{dI}{dV}$.

\section{Floquet Generation of higher-order topological superconductor}\label{Sec:VI}
\begin{figure}[]
	\centering
	\subfigure{\includegraphics[width=0.48\textwidth]{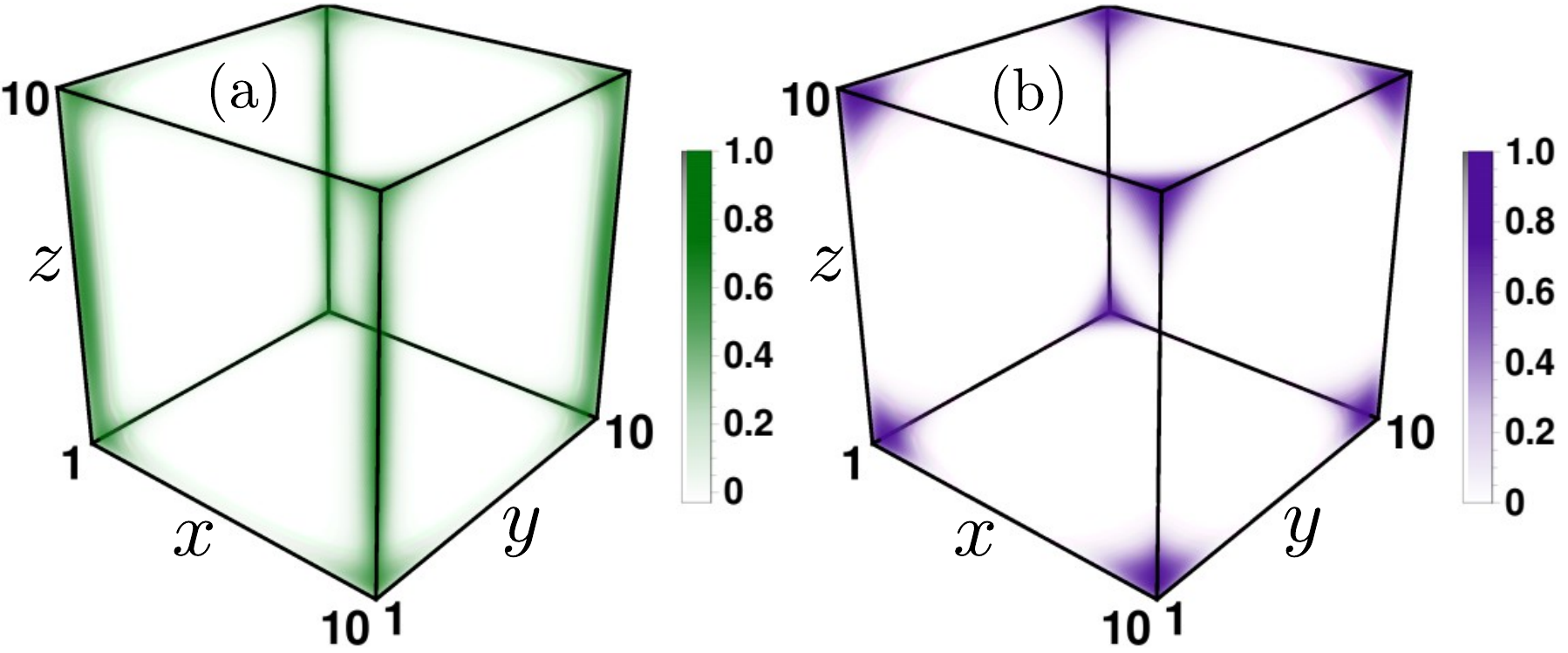}}
	\caption{(Color online) (a)~The structure of LDOS for the quasi-energy states corresponding to $\epsilon_m=0$, under OBC in all directions, is depicted while 
		considering the driving protocol (Eq.(\ref{kick})) with $\Lambda_1=0.4$ and $\Lambda_2 = 0.0$. 
		The clear signature of MHMs in FSOTSC, propagating along $z$-direction, can be observed at the interfaces of $xz$ and $yz$ surface. 
		(b)~We repeat (a) for $\Lambda_1=0.4$ and $\Lambda_2 = 0.4$ where MCMs are clearly visible referring to a FTOTSC phase. We choose the other parameter values as 
		$m=2.0, t=\lambda=1.0, \Delta=0.4, T=0.628$.
	}
	\label{FloquetLDOS}
\end{figure}
Having investigated the hierarchy of static HOTSC phases, we intend to discuss the Floquet generation of  
HOTSC phases starting from a 3D TI with proximitized $s$-wave superconductivity in it. We incorporate the following periodic kick driving protocol as~ \cite{Nag19,ghosh2020floquet,ghosh2020floquet2,Nag2020}
\begin{eqnarray}
	V(t)&=& {\tilde {\bm M}}({\bm k}) \cdot{\tilde {\bm \Gamma}}   \sum_{r=1}^{\infty} \delta(\tilde{t}-rT),
	\label{kick}
\end{eqnarray}
with the convention ${\tilde{\bm M}}({\bm k})= (V_1, V_2)$ and ${\tilde {\bm \Gamma}}= (\Gamma_6,\Gamma_7)$ where $T$ denotes the period of the drive and $\tilde{t}$ represents time. Similar to the static case, $V_1 \ne 0 $, and $V_2 =0$ 
($V_1 \ne 0$, and $V_2 \ne 0$)  engender FSOTSC (FTOTSC).  Using the static Hamiltonian (Eq.(\ref{Ham})) with $V_1=V_2=h_x=0$, the Floquet operator reads as  
$U(T)=\exp(-i H(\bm k) T)~\exp(-i V)$. Thus, one can obtain the effective Floquet Hamiltonian which is valid at any frequency and can be written as,  
\small
\begin{eqnarray}
	H_{\rm Eff}&=&\frac{\xi_{\bm k}}{\sin\xi_{\bm k}T}\Bigg[\sin(\left|{{\bm N}}({\bm k})\right| T) \cos(\left|{{ \bm M}}({\bm k})\right|) \sum_{j=1}^{5}n_j \Gamma_j \non \\ &&+\cos(\left|{{\bm N}}({\bm k})\right| T) \sin(\left|{{\bm M}}({\bm k})\right|) \sum_{j=1,2} m_j \Gamma_{j+5}\non \\
	&&+\sin(\left|{{\bm N}}({\bm k})\right| T) \sin(\left|{{\bm M}}({\bm k})\right|) \sum_{j=1}^{5} \left(n_j m_1 \Gamma_{j6} + n_j m_2 \Gamma_{j7}\right) \Bigg]\ , \non \\
\end{eqnarray}
\vskip -0.2cm
\normalsize
with $\xi_{{\bm k}}=\frac{1}{T}\cos^{-1} \left[\cos(\left|{{\bm N}}({\bm k})\right| T) \cos(\left|{{\bm M}}({\bm k})\right|) \right]$, $n_j=\frac{N_j({\bm k})}{\left|{{\bm N}}({\bm k})\right|}$, $m_j=\frac{M_j({\bm k})}{\left|{{\bm M}}({\bm k})\right|}$, 
$|{\bm N} ({\bm k}) |=\sqrt{ \sum^5_{j=1} N^2_j({\bm k})}$ and $|{\bm M} ({\bm k}) |=\sqrt{ \sum^2_{j=1} M^2_j({\bm k})}$. 

We numerically solve the Floquet operator $U(T)\ket{\phi_m}=\exp(-i\epsilon_m T)\ket{\phi_m}$ to obtain quasi-energy states $\ket{\phi_m}$ corresponding to the quasienergy 
$\epsilon_m$. 
We demonstrate the LDOS, associated with the zero (within numerical accuracy $\epsilon_m \simeq 0$) quasi-energy states, in Figs.~\ref{FloquetLDOS}(a) and (b) for 1D Floquet MHMs and 0D Floquet MCMs respectively with appropriate driving parameters. It is worth mentioning that these Floquet MHMs and MCMs are protected by the PHS $\mathcal{C}$. For the topological characterization of these 
FSOTSC and FTOTSC phases, one can make resort to Floquet WS that exhibits mid-gap eigenvalues at $0.5$, similar to the static case~\cite{ghosh2020floquet,ghosh2020floquet2}. 
For the sake of simplicity, we restrict ourselves to the case with no magnetic field \ie $h_x=0$~\cite{future_study}. Note that the FHOTSC phases, obtained here in the high frequency regime
adopting dynamical protocol (Eq.(\ref{kick})), do not conceive anomalous modes with quasienergy $\epsilon_m = \pi/T$~\cite{ghosh2020floquet,ghosh2020floquet2}. Therefore, the generation of 
anomalous HOTSC modes (dynamical MHMs or MCMs at quasienergy $\frac{\pi}{T}$), via appropriate Floquet driving~\cite{Huang2020,WuPRBL2021} is still an open question and will be presented elsewhere.

\section{Discussions and Summary}\label{Sec:VII}

To summarize, in this article, we propose a fermionic model based on 3D TI with proximity induced $s$-wave superconductivity to realize 
both SOTSC and TOTSC hosting 1D MHMs and 0D MCMs respectively, under the application of appropriate Wilson-Dirac mass  perturbations. The low energy effective model allows us to verify the 
above numerical observations analytically. Interestingly, application of a finite magnetic field in the bulk permits one to manipulate the number of MHMs leaving the MCMs unaltered. We characterize 
these topological phases by distinct distribution of WS. We also illustrate the quantized transport signature of MHMs in a two-terminal setup. Finally, we demonstrate a prescription to generate the FSOTSC 
and FTOTSC phases by periodically kicking the static Wilson-Dirac mass perturbations. 

As far as possible experimental feasibility of our model is concerned, superconductivity can be induced in 3D TI surface states ($\rm Bi_{2}Se_{3}$, HgTe etc.) via the proximity effect~\cite{Experiment3DTIProximity1,PhysRevB.96.165302,Experiment3DTIProximity2} with an induced gap $\Delta_{0}\sim ~\rm 0.5~meV$~\cite{Experiment3DTIProximity2}. 
The Wilson-Dirac mass perturbations may in principle be realized in optical lattice platform~\cite{huang2016experimental,Eckardt2017}. 
In recent times, the hierarchy of HOT phases in 3D has been experimentally discovered in sonic crystals~\cite{Experiment3DHOTI.aSonicCrystals}. 
Very recently, evidence of a HOTI in 3D has been reported
in van der Waals stacking of $\rm Bi_{4}Br_{4}$ chains~\cite{Experiment3DHOTI.VanDerWaals} via angle-resolved photoemission spectroscopy measurements.  
Given the experimental progress in this research field, we believe that our theoretical model proposal for MHMs and MCMs is timely and may be possible to realize in future experiments. 
However, the exact description of experimental techniques and prediction of candidate material are not the subject matter of our present manuscript. 

\subsection*{Acknowledgments}
 AKG and AS acknowledge SAMKHYA: High-Performance Computing Facility provided by Institute of Physics, Bhubaneswar, for numerical computations. TN acknowledges Bitan Roy and Vladimir 
 Juri\v ci\' c for stimulating discussions.




\appendix

\newcounter{defcounter}
\setcounter{defcounter}{0}




\vspace{20pt}

	\section{Low Energy Surface Theory}{\label{App:A}}
	We begin by writing down the Hamiltonian (Eq.(2) in the main text) around $\Gamma=(0,0,0)$ point as
	
	\begin{eqnarray}\label{HamLow}
	H_{\Gamma}&=& 2 \lambda \sum_{j=1}^{3} k_j \Gamma_j+\left(m_0-t  \sum_{j=1}^{3} k_j^2 \right) \Gamma_4 +\Delta_0 \Gamma_5 \non \\
	&&-\frac{\sqrt{3}\Lambda_1}{2} \left(k_x^2-k_y^2\right) \Gamma_6 - \frac{\Lambda_2}{2} \left(2k_z^2-k_x^2-k_y^2\right)  \Gamma_7 \non \\
	&&+h_x \Gamma_8 \ ,
	\end{eqnarray}	
	with the convention $k_{1,2,3}=k_{x,y,z}$ and $\Gamma_1=\mu_x\sigma_x s_x \tau_z$, $\Gamma_2=\mu_x \sigma_x s_y\tau_z$,  $\Gamma_3=\mu_x\sigma_x s_z \tau_z$, $ \Gamma_4=\mu_x\sigma_z \tau_z$, $\Gamma_5=  \tau_x$, $\Gamma_6=\mu_x \sigma_y$, $\Gamma_7=\mu_z \tau_z$ and $\Gamma_8=s_x$.  
	Here, the Pauli matrices ${\bm \mu}$, ${\bm \sigma}$, ${\bm s}$, and ${\bm \tau}$ act on sub-lattice $(A,B)$, orbital $(\alpha,\beta)$, spin $(\uparrow, \downarrow)$, and particle-hole ({\it e-h})   degrees of freedom respectively.

	\subsection{$xy$ surface}
	To derive the surface Hamiltonian for $xy$ surface, we consider open boundary condition (OBC) in the $z$ direction and periodic boundary condition (PBC) along $x$ and $y$ directions. The low-energy Hamiltonian (Eq.(\ref{HamLow})), can thus be written in two parts by replacing $k_z \rightarrow - i \partial_z $ and keeping upto first-order terms for $k_x$ and $k_y$, as~\cite{ghosh2020floquet,ghosh2020floquet2}: 
	\begin{eqnarray}\label{HLowxy}
	H_{\rm I} &=& \left(m_0+t \ \partial_z^2 \right) \Gamma_4 - 2 i \lambda \partial_z \Gamma_3\ , \non \\
	H_{\rm II}&=&2\lambda k_x \Gamma_1 +2\lambda k_y \Gamma_2 + \Delta \Gamma_5+\Lambda_2 \partial_z^2 \ \Gamma_7+h_x \Gamma_8\ . \qquad
	\end{eqnarray}
	Now one can solve for $H_{\rm I} |\Psi\rangle = 0$ considering the boundary condition $\ket{\Psi} \rightarrow 0$ as $z \rightarrow 0 , \infty$. Thus, we obtain 
	\begin{eqnarray}
	\ket{\Psi}=\mathcal{A} e^{-K_1 z} \sin K_2z \ e^{i k_x x+ik_y y} \ket{\chi} \ ,
	\end{eqnarray}
	where, $K_1=\frac{\lambda}{t}$, $K_2=\sqrt{\frac{m}{t}-\frac{\lambda^2}{t^2}}$, $\lvert \mathcal{A} \rvert^2= \frac{4 K_1 \left(K_1^2+K_2^2\right)}{K_2^2}$ and $\ket{\chi}$ is a $16$-component spinor  satisfying $\sigma_y s_z \ket{\chi}=+\ket{\chi}$. The latter can be choosen as follows, 
	\begin{eqnarray}
	\ket{\chi_1}&=&\ket{\mu_z=+1}\otimes\ket{\sigma_y=+1}\otimes\ket{s_z=+1}\otimes\ket{\tau_z=+1},\non\\
	\ket{\chi_2}&=&\ket{\mu_z=+1}\otimes\ket{\sigma_y=+1}\otimes\ket{s_z=+1}\otimes\ket{\tau_z=-1},\non\\
	\ket{\chi_3}&=&\ket{\mu_z=+1}\otimes\ket{\sigma_y=-1}\otimes\ket{s_z=-1}\otimes\ket{\tau_z=+1},\non\\
	\ket{\chi_4}&=&\ket{\mu_z=+1}\otimes\ket{\sigma_y=-1}\otimes\ket{s_z=-1}\otimes\ket{\tau_z=-1},\non\\
	\ket{\chi_5}&=&\ket{\mu_z=-1}\otimes\ket{\sigma_y=+1}\otimes\ket{s_z=+1}\otimes\ket{\tau_z=+1},\non\\
	\ket{\chi_6}&=&\ket{\mu_z=-1}\otimes\ket{\sigma_y=+1}\otimes\ket{s_z=+1}\otimes\ket{\tau_z=-1},\non\\
	\ket{\chi_7}&=&\ket{\mu_z=-1}\otimes\ket{\sigma_y=-1}\otimes\ket{s_z=-1}\otimes\ket{\tau_z=+1},\non\\
	\ket{\chi_8}&=&\ket{\mu_z=-1}\otimes\ket{\sigma_y=-1}\otimes\ket{s_z=-1}\otimes\ket{\tau_z=-1}. \non \\
	\end{eqnarray}
	The matrix element of $H_{II}$ in this basis reads
	\begin{equation}
	H_{xy,\alpha\beta}^{\rm S}=\int_{0}^{\infty} dz \ \bra{\Psi_{\alpha}} H_{II} \ket{\Psi_{\beta}}\ ,
	\end{equation}
	with $\alpha,~\beta=1,~\cdots,~8$. 
	Thus, the corresponding Hamiltonian for the $xy$ surface is given by
	\begin{eqnarray}\label{xysurface}
	H^{\rm S}_{xy}&=&-2 \lambda k_x \sigma_x s_y \tau_z +2 \lambda k_y \sigma_x s_x \tau_z + M_{\Delta}  \tau_x - 2M_{\Lambda_2} \sigma_z \tau_z. \non \\
	\end{eqnarray}
	
	\subsection{$yz$ surface}
	To obtain the surface Hamiltonian for $yz$ surface, we invoke OBC along $x$ direction, while other two directions continue to obey PBC. We can proceed as before and find the corresponding 
	zero-energy state $\ket{\Psi}$ as 
	\begin{eqnarray}
	\ket{\Psi}=\mathcal{A} e^{-K_1 x} \sin K_2x \ e^{i k_y y+ik_z z} \ket{\xi} \ ,
	\end{eqnarray}
	where, $\ket{\xi}$ is a $16$-component spinor satisfying $\sigma_y s_x \ket{\xi}=+\ket{\xi}$ and our choosen basis reads
	
	\begin{eqnarray}
	\ket{\xi_1}&=&\ket{\mu_z=+1}\otimes\ket{\sigma_y=+1}\otimes\ket{s_x=+1}\otimes\ket{\tau_z=+1},\non\\
	\ket{\xi_2}&=&\ket{\mu_z=+1}\otimes\ket{\sigma_y=+1}\otimes\ket{s_x=+1}\otimes\ket{\tau_z=-1},\non\\
	\ket{\xi_3}&=&\ket{\mu_z=+1}\otimes\ket{\sigma_y=-1}\otimes\ket{s_x=-1}\otimes\ket{\tau_z=+1},\non\\
	\ket{\xi_4}&=&\ket{\mu_z=+1}\otimes\ket{\sigma_y=-1}\otimes\ket{s_x=-1}\otimes\ket{\tau_z=-1},\non\\
	\ket{\xi_5}&=&\ket{\mu_z=-1}\otimes\ket{\sigma_y=+1}\otimes\ket{s_x=+1}\otimes\ket{\tau_z=+1},\non\\
	\ket{\xi_6}&=&\ket{\mu_z=-1}\otimes\ket{\sigma_y=+1}\otimes\ket{s_x=+1}\otimes\ket{\tau_z=-1},\non\\
	\ket{\xi_7}&=&\ket{\mu_z=-1}\otimes\ket{\sigma_y=-1}\otimes\ket{s_x=-1}\otimes\ket{\tau_z=+1},\non\\
	\ket{\xi_8}&=&\ket{\mu_z=-1}\otimes\ket{\sigma_y=-1}\otimes\ket{s_x=-1}\otimes\ket{\tau_z=-1}. \non \\
	\end{eqnarray}
	In this basis, we obtain the surface Hamiltonian for the $yz$ surface as 
	\begin{eqnarray}\label{yzsurface}
	H^{\rm S}_{yz}&=&2 \lambda k_y \sigma_x s_x \tau_z  +2 \lambda k_z \sigma_x s_y \tau_z + M_{\Delta}  \tau_x -M_{\Lambda_1} \sigma_x s_z   \non \\
	&&+h_x s_z + M_{\Lambda_2} \sigma_z \tau_z.
	\end{eqnarray}
	
	\subsection{$xz$ surface}
	Similarly for the $xz$ surface, we employ OBC along $y$ direction, while other two directions continue to obey PBC. The zero-energy state $\ket{\Psi}$ in this scenario can be written as 
	\begin{eqnarray}
	\ket{\Psi}=\mathcal{A} e^{-K_1 y} \sin K_2y \ e^{i k_x x +ik_z z} \ket{\zeta} \ ,
	\end{eqnarray}
	where, $\ket{\zeta}$ is a $16$-component spinor satisfying $\sigma_y s_y \ket{\zeta}=+\ket{\zeta}$ and our choosen basis reads
	
	\begin{eqnarray}
	\ket{\zeta_1}&=&\ket{\mu_z=+1}\otimes\ket{\sigma_y=+1}\otimes\ket{s_y=+1}\otimes\ket{\tau_z=+1},\non\\
	\ket{\zeta_2}&=&\ket{\mu_z=+1}\otimes\ket{\sigma_y=+1}\otimes\ket{s_y=+1}\otimes\ket{\tau_z=-1},\non\\
	\ket{\zeta_3}&=&\ket{\mu_z=+1}\otimes\ket{\sigma_y=-1}\otimes\ket{s_y=-1}\otimes\ket{\tau_z=+1},\non\\
	\ket{\zeta_4}&=&\ket{\mu_z=+1}\otimes\ket{\sigma_y=-1}\otimes\ket{s_y=-1}\otimes\ket{\tau_z=-1},\non\\
	\ket{\zeta_5}&=&\ket{\mu_z=-1}\otimes\ket{\sigma_y=+1}\otimes\ket{s_y=+1}\otimes\ket{\tau_z=+1},\non\\
	\ket{\zeta_6}&=&\ket{\mu_z=-1}\otimes\ket{\sigma_y=+1}\otimes\ket{s_y=+1}\otimes\ket{\tau_z=-1},\non\\
	\ket{\zeta_7}&=&\ket{\mu_z=-1}\otimes\ket{\sigma_y=-1}\otimes\ket{s_y=-1}\otimes\ket{\tau_z=+1},\non\\
	\ket{\zeta_8}&=&\ket{\mu_z=-1}\otimes\ket{\sigma_y=-1}\otimes\ket{s_y=-1}\otimes\ket{\tau_z=-1}. \non \\
	\end{eqnarray}
	We obtain the surface Hamiltonian for the $xz$ surface, in this basis, as 
	\begin{eqnarray}\label{xzsurface}
	H^{\rm S}_{xz}&=&-2 \lambda k_x \sigma_x s_x \tau_z  +2 \lambda k_z \sigma_x s_y \tau_z + M_{\Delta} \tau_x +M_{\Lambda_1} \sigma_x s_z   \non \\
	&&+ M_{\Lambda_2} \sigma_z \tau_z.
	\end{eqnarray}

	\section{Low Energy Hinge Theory}{\label{App:B}}
	
	In this section, we provide the Hamiltonian for different hinges propagating along $x$, $y$ and $z$ directions. We obtain a total of six hinge Hamiltonian from $xy$, $yz$ and $xz$ surface, depicted schematically in Fig.~\ref{HingeSchematics}.
	
	\subsection{Hinge from $xy$ surface}
	\subsubsection{Hinge along $x$ direction}
	To obtain the hinge Hamiltonian along $x$ direction from the $xy$ surface, we divide the surface Hamiltonian for $xy$ surface (Eq.(\ref{xysurface})) into two parts. Then considering OBC along 
	$y$ direction and PBC along $x$ direction one can write, 
	
	\begin{eqnarray}
	H_{\rm I}^{\rm S}&=&-2 i \lambda \partial_y \sigma_x s_x \tau_z + M_{\Delta} \tau_x, \non \\
	H_{\rm II}^{\rm S}&=&-\lambda k_x \sigma_x s_y \tau_z - 2 M_{\Lambda_2} \sigma_z\tau_z.
	\end{eqnarray}
	We solve for $H_{\rm I}^{\rm S} \ket{\Psi^{\rm S}} = 0$, considering the boundary condition  $\ket{\Psi^{\rm S}} \rightarrow 0$ as $y \rightarrow 0$. We obtain
	\begin{eqnarray}
	\ket{\Psi^{\rm S}_\alpha} \sim e^{-\xi_\alpha y+i k_x x} \ket{\chi_\alpha},
	\end{eqnarray}
	with $\xi_\alpha=\left\{\frac{M_{\Delta}}{2 \lambda},\frac{M_{\Delta}}{2 \lambda},\frac{M_{\Delta}}{2 \lambda},\frac{M_{\Delta}}{2 \lambda}\right\}$ and $\ket{\chi_\alpha}$ is given as
	
	\begin{eqnarray}
	\ket{\chi_\alpha}=\begin{pmatrix*}[r]
	-i & -i & -i & -i \\
	1 & -1 & 1 & -1 \\
	-i & i & i & -i \\
	1 & 1 & -1 & -1 \\
	-i & -i & i & i \\
	1 & -1 & -1 & 1 \\
	-i & i & -i & i \\
	1 & 1 & 1 & 1 
	\end{pmatrix*}\ .
	\end{eqnarray}
	
	The Hamiltonian for the hinge along $x$ direction can be obtained by calculating the matrix element of $H_{\rm II}^{\rm S}$ in $\ket{\Psi^{\rm S}_\alpha}$, as 
	\begin{equation}
	H_{x,xy}= 2 \lambda k_x \tau_y - 2 M_{\Lambda_2} s_x \tau_x\ ,
	\end{equation}
	where hinge is gapped due to the second perturnation $\Lambda_{2}$.
	
	\subsubsection{Hinge along $y$ direction}
	To derive the hinge Hamiltonian along $y$ direction, we consider OBC along $x$ direction and PBC along $y$ direction. Thus, we obtain the zero-energy solution as 
	\begin{eqnarray}
	\ket{\Psi^{\rm S}_\alpha} \sim e^{-\xi_\alpha x+i k_y y} \ket{\zeta_\alpha},
	\end{eqnarray}
	with $\xi_\alpha=\left\{\frac{M_{\Delta}}{2 \lambda},\frac{M_{\Delta}}{2 \lambda},\frac{M_{\Delta}}{2 \lambda},\frac{M_{\Delta}}{2 \lambda}\right\}$ and $\ket{\zeta_\alpha}$ is given as
	\begin{eqnarray}
	\ket{\zeta_\alpha}=\begin{pmatrix*}[r]
	-1 & -1 & -1 & -1 \\
	i & -i & i & -i \\
	i & -i & -i & i \\
	1 & 1 & -1 & -1 \\
	-1 & -1 & 1 & 1 \\
	i & -i & -i & i \\
	i & -i & i & -i \\
	1 & 1 & 1 & 1 
	\end{pmatrix*}.
	\end{eqnarray}
	The Hamiltonian for the hinge along $y$ direction is obtained as 
	\begin{equation}
	H_{y,xy}= -2 \lambda k_y \tau_y - 2 M_{\Lambda_2} s_x \tau_x\ .
	\end{equation}
	
	\subsection{Hinge from $yz$ surface}
	\subsubsection{Hinge along $y$ direction}
	For hinge Hamiltonian along $y$ direction we consider OBC along $z$ and PBC along $y$ direction respectively. The two parts of the $yz$ surface Hamiltonian (Eq.(\ref{yzsurface})) 
	can be written as 
	\begin{eqnarray}
	H_{\rm I}^{\rm S}&=&-2 i \lambda \partial_y \sigma_x s_x \tau_z + M_{\Delta} \tau_x -M_{\Lambda_1} \sigma_x s_z, \non \\
	H_{\rm II}^{\rm S}&=& 2\lambda k_z \sigma_x s_y \tau_z +  M_{\Lambda_2} \sigma_z\tau_z + h_x s_z.
	\end{eqnarray}
	
	The zero-energy solution is obtained as 
	\begin{eqnarray}
	\ket{\Psi^{\rm S}_\alpha} \sim e^{-\xi_\alpha z+i k_y y} \ket{\chi_\alpha},
	\end{eqnarray}
	with $\xi_\alpha=\left\{\frac{M_{\Lambda_1}+M_{\Delta}}{2 \lambda},\frac{M_{\Lambda_1}+M_{\Delta}}{2 \lambda},\frac{M_{\Lambda_1}-M_{\Delta}}{2 \lambda},\frac{M_{\Lambda_1}-M_{\Delta}}{2 \lambda}\right\}$ and $\ket{\chi_\alpha}$ is given as-
	\begin{eqnarray}
	\ket{\chi_\alpha}=\begin{pmatrix*}[r]
	1 & 1 & 1 & 1 \\
	1 & -1 & 1 & -1 \\
	1 & 1 & 1 & 1 \\
	-1 & 1 & -1 & 1 \\
	-1 & 1 & 1 & -1 \\
	-1 & -1 & 1 & 1 \\
	-1 & 1 & 1 & -1 \\
	1 & 1 & -1 & -1 
	\end{pmatrix*}.
	\end{eqnarray}
	The Hamiltonian for the hinge along $y$ direction is obtained as 
	\begin{equation}
	H_{y,yz}= -2 \lambda k_y s_z \tau_z +  M_{\Lambda_2} \tau_x\ .
	\end{equation}
	
	\subsubsection{Hinge along $z$ direction}
	For hinge Hamiltonian along $z$ direction we consider OBC along $y$ direction and PBC along $z$ direction, and obtain the zero-energy solution as 
	\begin{eqnarray}
	\ket{\Psi^{\rm S}_\alpha} \sim e^{-\xi_\alpha y+i k_z z} \ket{\zeta_\alpha}\ ,
	\end{eqnarray}
	with $\xi_\alpha=\left\{\frac{M_{\Lambda_1}+M_{\Delta}}{2 \lambda},\frac{M_{\Lambda_1}+M_{\Delta}}{2 \lambda},\frac{M_{\Lambda_1}-M_{\Delta}}{2 \lambda},\frac{M_{\Lambda_1}-M_{\Delta}}{2 \lambda}\right\}$ and $\ket{\zeta_\alpha}$ is given as-
	\begin{eqnarray}
	\ket{\zeta_\alpha}=\begin{pmatrix*}[r]
	1 & 1 & 1 & 1 \\
	1 & -1 & 1 & -1 \\
	-i & -i & -i & -i \\
	i & -i & i & -i \\
	-1 & 1 & 1 & -1 \\
	-1 & -1 & 1 & 1 \\
	i & -i & -i & i \\
	-i & -i & i & i 
	\end{pmatrix*}.
	\end{eqnarray}
	The Hamiltonian for the hinge propagating along $z$ direction can be obtained as 
	\begin{equation}
	H_{z,yz}= 2 \lambda k_z s_z \tau_z +  M_{\Lambda_2} \tau_x\ .
	\end{equation}
	\setcounter{figure}{0}
	\renewcommand{\thefigure}{B\arabic{figure}}
	\begin{figure}
		\centering
		\subfigure{\includegraphics[width=0.32\textwidth]{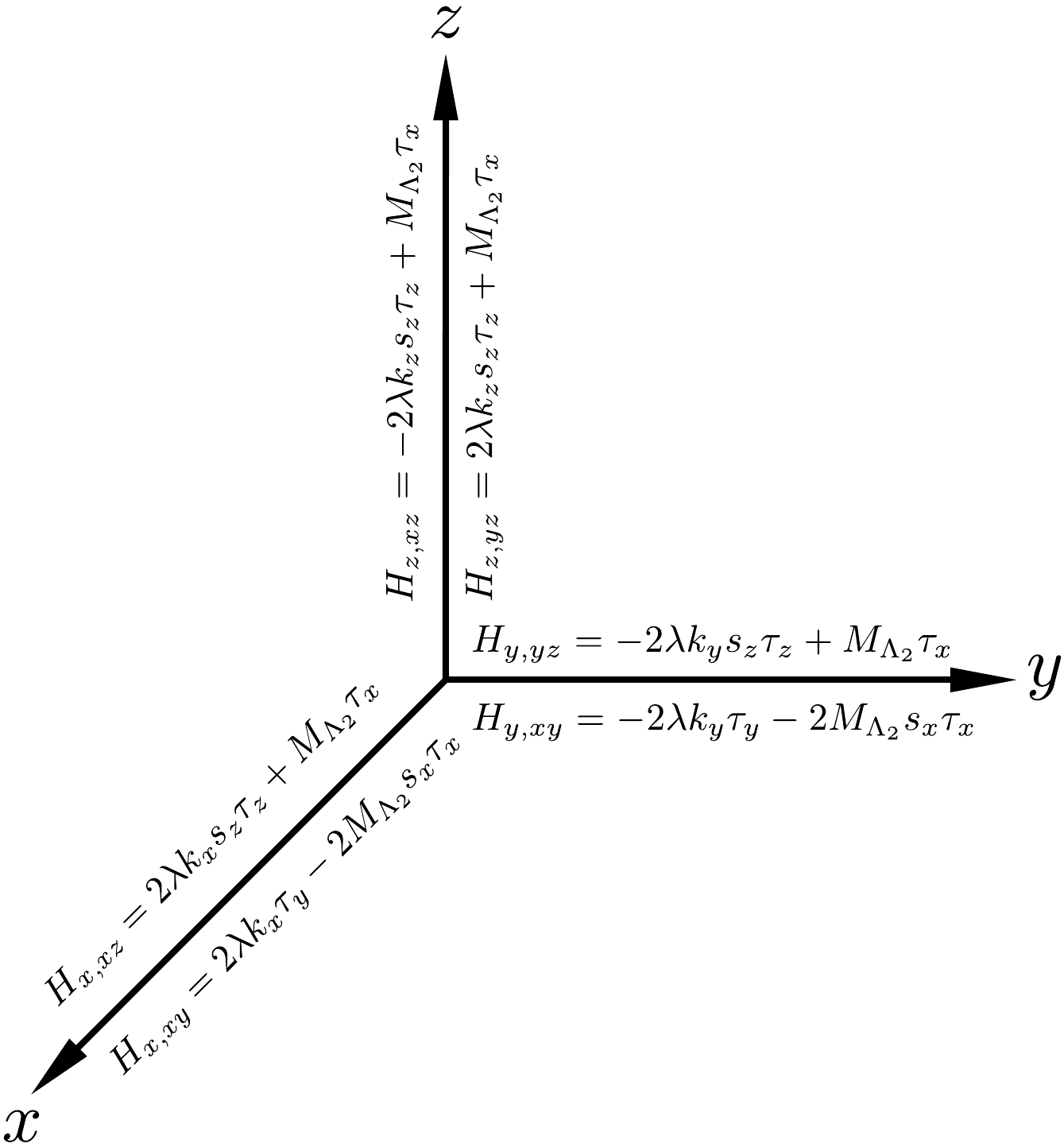}}
		\caption{Schematic diagram of various hinge Hamiltonians, obtained from different surfaces, is illustrated along $x$, $y$ and $z$ direction.}
		\label{HingeSchematics}
	\end{figure}
	
	\subsection{Hinge from $xz$ surface}
	
	\subsubsection{Hinge along $x$ direction}
	For hinge Hamiltonian along $x$ direction, we consider OBC along $z$ direction and PBC along $x$ direction. As before, the two parts of the surface Hamiltonian reads as 
	\begin{eqnarray}
	H_{\rm I}^{\rm S}&=& 2 i \lambda \partial_x \sigma_x s_x \tau_z + M_{\Delta} \tau_x +M_{\Lambda_1} \sigma_x s_z, \non \\
	H_{\rm II}^{\rm S}&=& 2\lambda k_z \sigma_x s_y \tau_z +  M_{\Lambda_2} \sigma_z\tau_z.
	\end{eqnarray}

	Then the zero-energy solution is obtained as 
	\begin{eqnarray}
	\ket{\Psi^{\rm S}_\alpha} \sim e^{-\xi_\alpha z+i k_x x} \ket{\chi_\alpha},
	\end{eqnarray}
	with $\xi_\alpha=\left\{\frac{M_{\Lambda_1}+M_{\Delta}}{2 \lambda},\frac{M_{\Lambda_1}+M_{\Delta}}{2 \lambda},\frac{M_{\Lambda_1}-M_{\Delta}}{2 \lambda},\frac{M_{\Lambda_1}-M_{\Delta}}{2 \lambda}\right\}$ and $\ket{\chi_\alpha}$ is given as
	\begin{eqnarray}
	\ket{\chi_\alpha}=\begin{pmatrix*}[r]
	1 & 1 & 1 & 1 \\
	1 & -1 & 1 & -1 \\
	-1 & -1 & -1 & -1 \\
	1 & -1 & 1 & -1 \\
	1 & -1 & -1 & 1 \\
	1 & 1 & -1 & -1 \\
	-1 & 1 & 1 & -1 \\
	1 & 1 & -1 & -1 
	\end{pmatrix*}.
	\end{eqnarray}
	The Hamiltonian for the hinge along $y$ direction can be written as 

	\begin{equation}
	H_{x,xz}= 2 \lambda k_x s_z \tau_z +  M_{\Lambda_2} \tau_x\ .
	\end{equation}
	
	\subsubsection{Hinge along $z$ direction}
	
	For hinge Hamiltonian along $z$ direction we employ OBC along $x$ direction and PBC along $z$ direction. Similar as before, we obtain the zero-energy solution as 
	\begin{eqnarray}
	\ket{\Psi^{\rm S}_\alpha} \sim e^{-\xi_\alpha x+i k_z z} \ket{\zeta_\alpha},
	\end{eqnarray}
	with
	
 $\xi_\alpha=\left\{\frac{M_{\Lambda_1}+M_{\Delta}}{2 \lambda},\frac{M_{\Lambda_1}+M_{\Delta}}{2 \lambda},\frac{M_{\Lambda_1}-M_{\Delta}}{2 \lambda},\frac{M_{\Lambda_1}-M_{\Delta}}{2 \lambda}\right\}$ and $\ket{\zeta_\alpha}$ is given as
 
	\begin{eqnarray}
	\ket{\zeta_\alpha}=\begin{pmatrix*}[r]
	1 & 1 & 1 & 1 \\
	1 & -1 & 1 & -1 \\
	-i & -i & -i & -i \\
	i & -i & i & -i \\
	1 & -1 & -1 & 1 \\
	1 & 1 & -1 & -1 \\
	-i & i & i & -i \\
	i & i & -i & -i 
	\end{pmatrix*}.
	\end{eqnarray}
	
	The Hamiltonian for the hinge along $z$ direction is obtained as 
	\begin{equation}
	H_{z,xz}= -2 \lambda k_z s_z \tau_z +  M_{\Lambda_2} \tau_x\ .
	\end{equation}
	
	\section{Corner Mode Solutions}{\label{App:C}}
	Here we provide the solution for the MCMs located at $x=y=z=0$. To obtain the same we solve the hinge Hamiltonians derived before. Then we find the appropriate solutions 
	(respecting the boundary condition) therein and match them at $x=y=z=0$. Thus, we find the solution for the MCMs as

	\begin{eqnarray}
	\Phi &\sim& c_1^x \ \phi_1 \ e^{-\frac{M_{\Lambda_2}x}{\lambda}}+ \ c_2^x \ \phi_2 \ e^{-\frac{M_{\Lambda_2}x}{2\lambda}} {\rm : along } \ x \non \\
	&\sim& c_1^y \ \phi_1 \ e^{-\frac{M_{\Lambda_2}y}{2\lambda}}+ \ c_2^y \ \phi_2 \ e^{-\frac{M_{\Lambda_2}y}{\lambda}} {\rm : along } \ y \non \\
	&\sim& c_1^z \ \phi_1 \ e^{-\frac{M_{\Lambda_2}z}{2\lambda}}+ \ c_2^z \ \phi_2 \ e^{-\frac{M_{\Lambda_2}z}{2\lambda}} {\rm : along } \ z \ ,
	\label{corner} 
	\end{eqnarray}
	
	Here, $c_{1,2}^{x,y,z}$ are arbitrary constants, $\phi_1= \{1,i,-1,i\}^T$ and $\phi_2= \{1,-i,1,i\}^T$ are the spinors.
	
	
	\section{SOTI-SOTSC-SOTI junction}{\label{App:D}}

	In this section we discuss the schematic (see Fig.~\ref{TransportSchematics}) and outline of our transport set-up that we employ to compute the differential conductance, $\frac{dI}{dV}$ in order
	to obtain the transport signature of MHMs. 
	As mentioned earlier and also evident from Fig.~\ref{TransportSchematics} is that the leads we use are SOTI. Here we provide the details of our lattice model used in KWANT~\cite{groth2014kwant} 
	to calculate $\frac{dI}{dV}$.  \\
	
	\begin{widetext}
		
		\begin{eqnarray}
		H&=& \sum_{x,y,z,\alpha,\beta}\Bigg[C^\dagger_{\alpha,x,y,z} \Big\{ (m_0-6 t) \Gamma_4^{\alpha\beta} +\Delta \Gamma_5^{\alpha\beta} +h_x\Gamma_8^{\alpha\beta} \Big\}  C_{\beta,x,y,z}
		+C^\dagger_{\alpha,x,y,z}  \Big\{ i \lambda \Gamma_1^{\alpha\beta} +t \Gamma_4^{\alpha\beta} +\frac{\sqrt{3}\Lambda_1}{2}\Gamma_6^{\alpha\beta} -\frac{\Lambda_2}{2}\Gamma_7^{\alpha\beta} \Big\}  C_{\beta,x+1,y,z}  \non \\
		&&+ C^\dagger_{\alpha,x,y,z}  \Big\{ i \lambda \Gamma_2^{\alpha\beta} +t \Gamma_4^{\alpha\beta} -\frac{\sqrt{3}\Lambda_1}{2}\Gamma_6^{\alpha\beta} -\frac{\Lambda_2}{2}\Gamma_7^{\alpha\beta} \Big\}  C_{\beta,x,y+1,z}
		+ C^\dagger_{\alpha,x,y,z}  \Big\{ i \lambda \Gamma_3^{\alpha\beta} +t \Gamma_4^{\alpha\beta}  +\Lambda_2\Gamma_7^{\alpha\beta} \Big\}  C_{\beta,x,y,z+1} + {\rm h.c.} \Bigg] \non, \\
		\label{real_space}
		\end{eqnarray}
	\end{widetext}

	Here, $\alpha, \beta$ index encapsulate all four \ie sub-lattice $(A,B)$, orbital $(\delta,\gamma)$, spin $(\uparrow, \downarrow)$, and particle-hole ({\it {e-h}}) degrees of freedom. $\Delta$ is the  
	superconducting order parameter which is taken to be zero for the SOTI leads \ie left and right regions, whereas $\Delta=\Delta_0$ for the central SOTSC region (see Fig.~\ref{TransportSchematics}). The momentum space version of Eq.(\ref{real_space}) is given by Eq.(\ref{Ham}). As the $dI/dV$ depends on the scattering probabilities (via Landauer-B\"{u}ttiker formula) through our 
	transport setup, the low energy effective model will yield the similar results as shown via our tight-binding lattice model (see Fig.~\ref{Wannier}(d)).

\bibliography{bibfile}{}
\end{document}